# An Initial Study on Load Forecasting Considering Economic Factors

Hossein Sangrody, *Student Member, IEEE*
Electrical and Computer Engineering Department
State University of New York, Binghamton University
Binghamton, NY 13902, USA
habdoll1@binghamton.edu

Ning Zhou, *Senior Member, IEEE*
Electrical and Computer Engineering Department
State University of New York, Binghamton University
Binghamton, NY 13902, USA
ningzhou@binghamton.edu

*Abstract*— This paper proposes a new objective function and quantile regression (QR) algorithm for load forecasting (LF). In LF, the positive forecasting errors often have different economic impact from the negative forecasting errors. Considering this difference, a new objective function is proposed to put different prices on the positive and negative forecasting errors. QR is used to find the optimal solution of the proposed objective function. Using normalized net energy load of New England network, the proposed method is compared with a time series method, the artificial neural network method, and the support vector machine method. The simulation results show that the proposed method is more effective in reducing the economic cost of the LF errors than the other three methods.

## I. INTRODUCTION

Load forecasting (LF) is a critical component in power system operation and planning. Depending on the purposes of LF, the lead times of LF can vary from seconds to years. Very short-term load forecasting (VSTLF) [1] and short-term load forecasting (STLF) [2] usually have lead times of seconds to weeks and are often used for control and, operation purposes. In contrast, medium-term load forecasting (MTLF) [3] and long term load forecasting (LTLF) [4] have lead times of month(s), years, even decades and are often used for scheduling and planning purposes.

Many methods and models have been proposed to forecast load. According to methodologies used, LF methods can be classified in three main categories: time series analysis, and machine learning, and hybrid methods. Time series analysis methods include multiple linear regression (MLR) [5], auto regressive moving average (ARMA) [6], autoregressive integrated moving average (ARIMA) [7], exponential smoothing [8]. Machine learning methods include artificial neural network (ANN) [9], fuzzy logic (FL) [10], support vector regression (SVR) [11]. In a hybrid method, two or more methods are combined together to improve forecasting accuracy. Because of the complexity of load behaviors, the effectiveness of different methods may vary for different application scenarios and evaluation metrics.

The popularity of hybrid methods has increased recently. In hybrid methods, the benefits of other methods, such as heuristic algorithms, FL, Kalman filter, are implemented in a LF model to improve forecasting efficiency [12]. ANN and regression methods or combination of them are the main methods applied in hybrid methods. Despite of the benefits of hybrid methods, their parameters need to be adjusted well to achieve accurate forecasting [12].

The mean absolute percentage error (MAPE) is a statistic metric, which is commonly used to evaluate the accuracy of LF methods. The MAPE gives relative errors in percentage, which does not depend on the scale of forecasted variables. Therefore, the MAPE has been widely used to compare forecasting accuracy under different scenarios. Also, by using absolute errors, the MAPE and some other metrics do not distinguish the direction of errors. In other words, the positive LF errors (i.e., when actual values > forecasted values) and negative LF errors (i.e., when actual values < forecasted values) are counted equally in the metrics.

Ignoring the direction of errors simplifies the efforts of evaluating forecasting accuracy. However, it should be noted that the directions of errors often have economic impact in some LF applications. For example, in LTLF for power system planning, positive LF errors (i.e. actual values > forecast values) can result in planning inadequate capacity and in turn loss of load service. On the other hand, negative LF errors (actual values < forecast values) can result in wasting of resources by deploying more capacity than necessary. Note that economic loss corresponding to losing load (due to positive errors) is often different from that corresponding to resource wasting (due to negative errors). Taking the difference into account, a new error metric and objective function with different economic coefficients for positive errors and negative errors is proposed in this paper. Quantile regression (QR) method is used to solve the defined problem. The proposed methodology is applied for the LTLF of the net energy load

(NEL) of New England network and the results are compared with three other LF methods.

The rest of the paper is organized as follows. In Section II, the LF problem is formulated into a new objective function. Section III proposes QR to solve the LF problem using a linear LF model. Section IV presents the simulation results. The conclusion is drawn in Section V.

## II. PROBLEM FORMULATION

The difference between the actual values and the forecasted values is called LF errors ($e_i$) and defined by (1).

$$e_i = y_i - f_\beta(x_{i-l}) \quad (1)$$

Here, $y_i$ is the actual value at time instant $i$. The symbol $x_{i-l}$ is the independent variable available at time instant $i-l$ for forecasting $y_i$. The symbol $l$ is for lead times. The $f_\beta$ is the forecast model, which is based on parameter vector of $\beta$. The function $f_\beta(x_{i-l})$ gives the forecasted value of $y_i$.

A commonly used metric for evaluating the accuracy of LF results is the MAPE defined by (2). Other metrics are mean absolute error (MAE) [13, 14] defined by (3), mean squared error (MSE) [15] defined by (4), and root-mean-square error (RMSE) [13, 14] defined by (5). Here, $N$ is the total number of time instants.

$$\text{MAPE} = \frac{1}{N}\sum_{i=0}^{N}\left|\frac{y_i - f_\beta(x_{i-l})}{y_i}\right| \times 100 \quad (2)$$

$$\text{MAE} = \frac{1}{N}\sum_{i=0}^{N}\left|y_i - f_\beta(x_{i-l})\right| \quad (3)$$

$$\text{MSE} = \frac{1}{N}\sum_{i=0}^{N}(y_i - f_\beta(x_{i-l}))^2 \quad (4)$$

$$\text{RMSE} = \sqrt{\frac{1}{N}\sum_{i=0}^{N}(y_i - f_\beta(x_{i-l}))^2} \quad (5)$$

Note that these metrics do not distinguish error directions. Yet, LF errors in different directions may have different economic impacts. As it was discussed in Section I, even though errors in both directions result into economic losses, the prices tagged to different error directions are often different. Accordingly, objective functions of LF methods should be able to attach different price tags to positive and negative LF errors.

Considering the different economic impact of positive and negative LF errors, we propose using (6) as an objective function for LF models. The objective function in (6) is named the economic load forecast error (ELFE). A LF model, which results in small ELFEs, is preferred.

$$\text{ELFE:} \quad J(\beta) = P_+ \sum_{i:y_i > f_\beta(x_{i-l})}^{N}\left|y_i - f_\beta(x_{i-l})\right|$$

$$+ P_- \sum_{i:y_i < f_\beta(x_{i-l})}^{N}\left|y_i - f_\beta(x_{i-l})\right| \quad (6)$$

In (6), the first term is for the total cost of the positive LF errors, which includes the time instants when $y_i > f_\beta(x_{i-l})$. The second term is for the total cost of negative LF errors, which includes the time instants when $y_i < f_\beta(x_{i-l})$. The symbol $P_+$ and $P_-$ are the prices attached to the positive and negative LF errors respectively, which account for financial cost of LF errors. Generally, in power systems, because the goal of a LF method is to forecast energy to be consumed by load, the unit of the price tags in (6) (i.e. $P_+$ and $P_-$) shall be dollar per energy unit. As such, the unit of the ELFE shall be dollar. Given $y_i$, $x_{i-l}$, $P_+$, and $P_-$, the optimal parameters of a forecast model (i.e. $\hat{\beta}$) can be determined using (7).

$$\hat{\beta} = argmin\{J(\beta)\} \quad (7)$$

## III. PROPOSED METHODOLOGY

This section proposes a linear LF model and applies quantile regression to obtain its optimal parameter. To prepare the problem for quantile regression, equation (8) is obtained by dividing the objective function in (6) by $(P_+ + P_-)$.

$$\frac{J(\beta)}{P_+ + P_-} = \frac{P_+}{P_+ + P_-}\sum_{i:y_i > f_\beta(x_{i-l})}^{N}\left|y_i - f_\beta(x_{i-l})\right|$$

$$+ \frac{P_-}{P_+ + P_-}\sum_{i:y_i < f_\beta(x_{i-l})}^{N}\left|y_i - f_\beta(x_{i-l})\right| \quad (8)$$

Assuming $\tau = \frac{P_+}{(P_+ + P_-)}$ and a linear LF model defined by (9), (8) can be written into (10). The optimal parameter of the LF model (9) can be found using (11).

$$f_\beta(x_{i-l}) = x_{i-l}^T \beta_\tau \quad (9)$$

$$\frac{J(\beta_\tau)}{P_+ + P_-} = \tau \sum_{i:y_i > x_{i-l}^T \beta_\tau}^{N}\left|y_i - x_{i-l}^T \beta_\tau\right|$$

$$+ (1-\tau) \sum_{i:y_i < x_{i-l}^T \beta}^{N}\left|y_i - x_{i-l}^T \beta_\tau\right| \quad (10)$$

$$\hat{\beta}_\tau = argmin\left\{\frac{J(\beta_\tau)}{P_+ + P_-}\right\} \quad (11)$$

Note that because $P_+$ and $P_-$ are both greater than zero, $\tau$ should be between 0 and 1. As such, equation (10) matches QR requirement. As a result, equations (10) and (11) can be solved for $\hat{\beta}_\tau$ using QR.

By modeling the relationship between inputs and conditional percentile of target values, QR assigns different penalties to the positive and negative errors, and minimizes the summation of errors [16]. Assuming $Y$ as a random dependent variable, $X$ as multidimensional independent variables, and $F_Y(y|X = x) = P(Y \leq y|X = x)$ as the conditional cumulative distribution function of $Y$ given $X = x$, conditional quantile, i.e. $Q_\tau$ and linear conditional quantile model are defined by (12) and (13), respectively.

$$Q_\tau(Y|X = x) = invf\{y : F_Y(y|x) > \tau\} \quad (12)$$

$$Q_\tau(Y|x) = x^T \beta_\tau, \ 0 < \tau < 1 \quad (13)$$

Here "*invf*" in (12) means the inverse function and $\beta_\tau$ is QR coefficient which is obtained by solving (14).

$$\hat{\beta}_\tau = argmin \sum_{i=1}^{N}(\tau - \mathbf{1}_{(y_i < x_i^T \beta_\tau)})(y_i - x_i^T \beta_\tau) \quad (14)$$

As a result, (14) is the representation of (10). Similar to the least squares regression, QR tries to get a solution that produces the smallest errors. Different from the least squares regression, QR punishes the errors instead of the squares of errors. Therefore, QR is more robust against outliers than the least squares regression. In addition, QR can attach different weights (i.e. $P_+$ and $P_-$) based on the direction of errors while the least squares regression does not distinguish the direction of errors. These advantages of QR come at a cost of computation complexity. While the least squares regression can be solved easily using linear projection, QR is a linear programming problem, which often takes longer time to solve.

## IV. SIMULATION RESULTS

To evaluate the performance of the proposed method, the normalized NEL of New England network shown in Fig.1 is used for case studies. The data set includes monthly normalized NEL from January 2000 to September 2015 along with historical temperature indicators including total heating degree day (HDD) and total cooling degree day (CDD) [17].

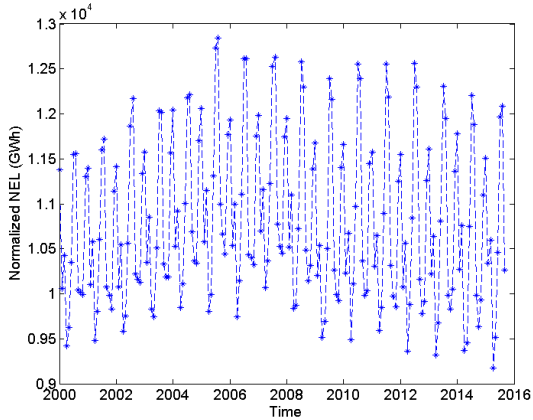

Figure 1. Normalized NEL in New England network from January 2000 to September 2015.

The normalized NEL is a monthly quantity, which is assigned as $y_i$ in (1). The lead time of the LF (i.e., $l$) is set one year. The independent variable $x_{i-l}$ includes HDD/CDD and the normalized NEL of the previous 11 years. Accordingly, to forecast monthly normalized NEL one-year ahead, the load data of the same month from past 11 years as well as HDD/CDD of the target month are used as input data. For example, to forecast the normalized NEL in January 2011, the load data of January 2000-2010 as well as total HDD and total CDD of the January in 2011 are used to the LF model as independent variables. Note that the actual HDD and CDD instead of their forecasted values are used to eliminate the potential influence of weather forecast models.

Different forecasting models can be used for $f_\beta(*)$. Because MLR, ANN, and SVR methods are pervasive in LF area, these methods along with QR are applied to forecast monthly normalized NEL from January 2011 to September 2015. These target data are divided into two groups: training data (with 60% of data) and validation data (with 40% of data).

The MLR and ANN methods for LF modeling are implemented using MATLAB®. For the ANN method, by trying with different hidden layers, we decide to use an ANN with one hidden layer and 10 neurons because it gives low forecasting errors for both training and validation data sets. Also, the Bayesian regularization is used for training. The SVR is applied for LF using LIBSVM [18]. The SVR type and kernel function are selected as nu-SVR and linear, respectively.

In this case study, we assume that deploying inadequate capacity results in more economic loss than deploying more-than-necessary capacity. That is $P_+ > P_-$ in (6). Accordingly, we consider percentile of 70% in (10) (i.e., $\tau = \mathbf{0.70}$). Note that depending on the prices assigned to LF errors (i.e., $P_+$ and $P_-$), $\tau$ may take different values that vary between 0 and 1.

The LF results of the QR method are shown in Fig. 2 for the time durations between January 2011 and September 2015. The data between 2011 and 2014 is used as training data and its forecast is shown as a blue dashed line with blue circle markers. The data between 2014 and 2015 is used as validation data and its forecast is shown as a red dashed line with red square markers. Observe that the forecasted values match well with the actual values for both the training and validation data. Metrics defined in (2), (3) and (5) are used to quantify the LF errors. The simulation results show that for the training data MAPE=0.74, MAE= 77.7, RMSE=119.7. For the validation data MAPE=1.99, MAE=211.3, RMSE=264.5. It can be observed that the errors for the validation data are larger than those for the training data.

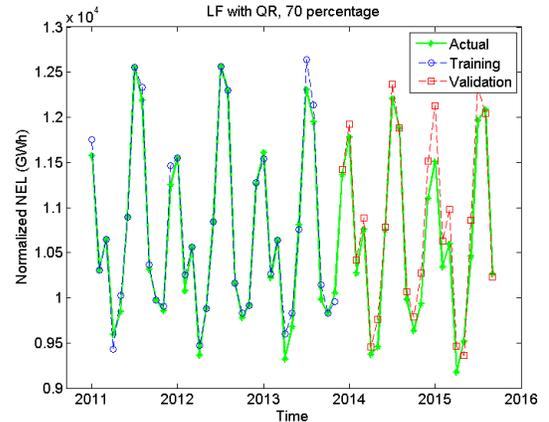

Figure 2. LF results of the normalized NEL of New England using the QR method (training data: MAPE=0.74, MAE=77.7, RMSE=119.7, ELFE=972 validation data: MAPE=1.99, MAE=211.25, RMSE=264.5, ELFE=1487).

The resulting MAPEs of all the LF methods are summarized in Table I. The small MAPEs in Table I indicate that all the methods can forecast load reasonably well. Note that because the ANN uses a random process in training its model, its corresponding MAPEs vary in different running of simulations.

Its MAPE in Table I is the average value of 50 times of running of the same ANN model.

TABLE I. LF RESULTS IN MAPE

| Method | MAPE for Training Data | MAPE for Validation Data |
|--------|------------------------|--------------------------|
| MLR | 0.66 | 1.72 |
| ANN | 0.81 | 1.78 |
| SVR | 1.07 | 2.12 |
| QR | 0.74 | 1.99 |

The MAPEs of all the methods using training and validation data are illustrated by the boxplots in Fig.3a and Fig.3b respectively. It can be observed that the validation data set results into larger MAPEs than the training data set for all the LF methods. In addition, the MAPEs of the four methods are all small and not significantly different from to each other.

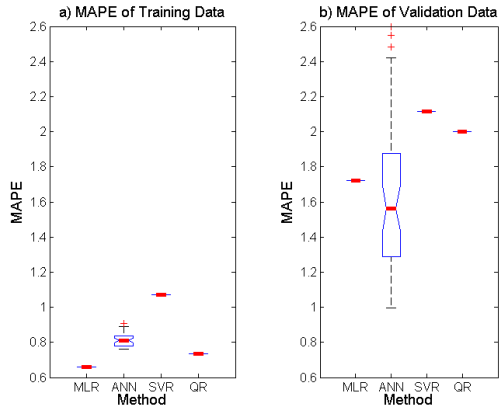

Figure 3. MAPEs for a) training data b) validation data.

Fig. 4 and Fig. 5 illustrate MAEs and RMSEs of the LF results from all the methods respectively. The left subplots of the figures are for the training data. The right subplots are for the validation data. Similar to the observation on the MAPE, the metrics in Fig. 4 and Fig. 5 indicate that all the LF methods produce reasonably good forecast. Also observe that relative LF accuracy may be different when we use different metrics. For example, for the training data, the MAPE and MAP of the QR are smaller than those of the ANN in Fig. 3.a and Fig. 4.a, while the RMSE of the QR is larger than that of the ANN in Fig. 5.a.

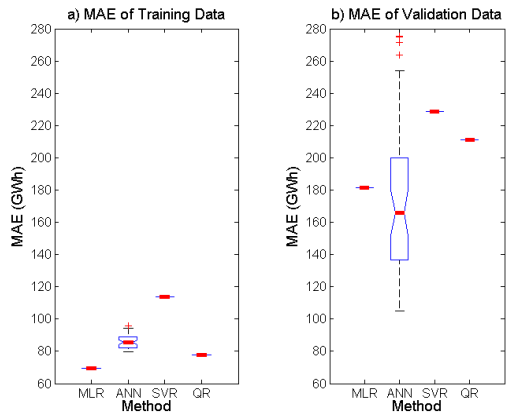

Figure 4. MAEs for a) training data b) validation data.

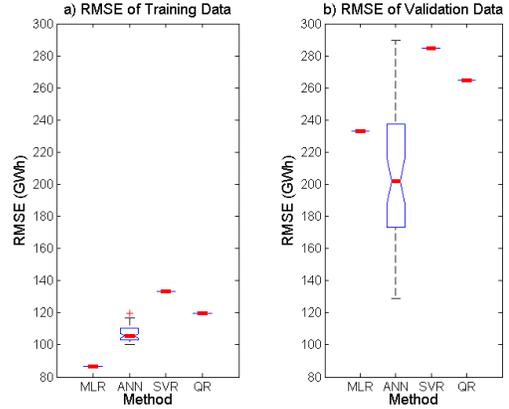

Figure 5. RMSEs a) training data b) validation data.

As it is indicated by (10), the performance of QR also depends on the value of $\tau$ and the load profile. The $\tau$ in QR is set 70% for the case study shown in Fig. 3 – Fig.5. To show the impact of $\tau$ on MAPE, $\tau$ is varied from 50% (the median regression) to 90% and the resulting MAPEs of the QR are shown in Fig. 6. Note that the MAPEs for the training data increases with the increase of $\tau$ because the MAPE is a metric that puts the same weights on positive and negative errors.

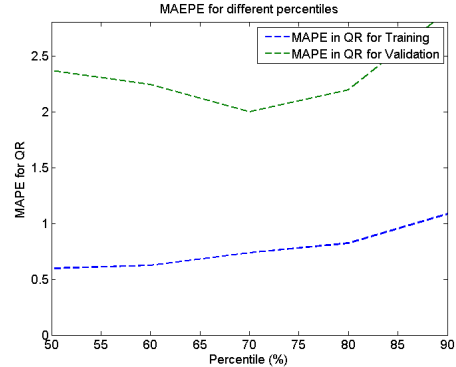

Figure 6. MAPEs for QR with different percentiles (i.e. $\tau$).

Table II shows the LF errors of all the methods using the proposed ELFE metric. Note that because in the case study, target value $y_i$ is energy ($GWh$), and price tag (i.e. $P_+$ and $P_-$) has the unit of dollar per energy ($\$/GWh$). Assuming that $(P_+ + P_-) = \$d/GWh$, the objective function of quantile regression is the ELFE divided by $d$ ($\$$). As shown by Table II, the ELFEs from the QR have the least values for both the training and validation data.

TABLE II. LF RESULTS WITH ELFE CRITERION

| Method | ELFE/d ($) for Training Data | ELFE/d ($) for Validation Data |
|--------|------------------------------|--------------------------------|
| MLR | 1214 | 2707 |
| ANN | 1500 | 2617 |
| SVR | 1914 | 1797 |
| QR | 972 | 1487 |

Similar to the previous setup, the simulation is run for 50 times for the ANN method because of the randomness in its training procedure. The average ELFE of the ANN over the whole 50 times of running is shown in Table II and the whole results of 50 times of running are illustrated in Fig.7a and Fig.7b for training and validation data respectively. Here the y-axis is ELFE/d ($).

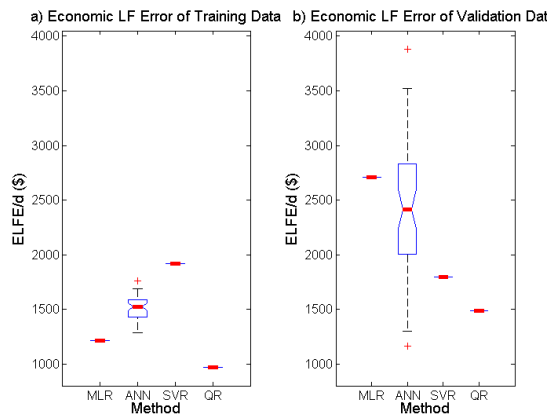

Figure 7. ELFE for forecasting the normalized NEL of New England. a) training data b) validation data.

Fig. 8 shows the ELFE for different percentiles of QR. As it is illustrated, the ELFEs decrease with the increase of percentiles. Depending on load forecast purposes and economic considerations, the percentile may take different values. In this case study, because we assume that $P_+ \geq P_-$ in (8), $\tau$ in (10) is varied between 50% and 90%. Fig. 8 illustrates the ELFEs with the scale of ELFE/d for different values of $\tau$ from 50% (median QR) to 90%.

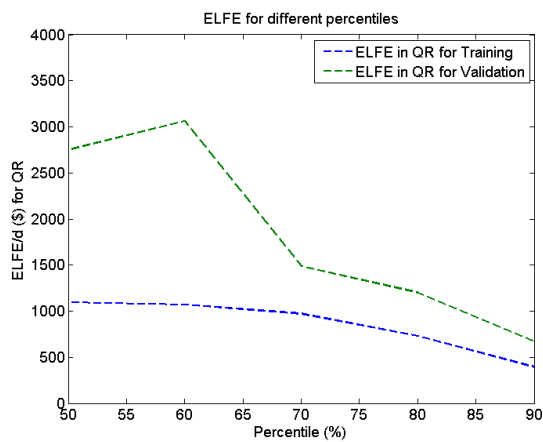

Figure 8. ELFEs for the QR method with different percentiles (i.e. $\tau$).

The objective function presented in this paper was applied for LTLF; Note that the objective function can also be applied for VSTLF, STLF, and MSTLF.

## V. CONCLUSION

In power systems, economic losses can be incurred by both positive and negative LF errors. Generally, the prices for different directions of LF error are often different. The MAPE and other commonly used metrics for evaluating LF accuracy do not distinguish the positive and negative LF errors. In this paper, a new metric (i.e., the ELFE) is proposed to attach different price tags to positive and negative LF errors in the objective function. The proposed problem is solved by QR method. The simulation using the normalized NEL from New England network shows that the proposed method is more effective in reducing the economic loss incurred by LF errors than the other commonly used LF methods when the price of positive LF errors is different from the price of negative LF errors.